\documentclass{article}

\usepackage{amsfonts}
\usepackage{graphicx}
\usepackage{amssymb}

\begin{document}

\title{A mechanical model of a PR-Box}
\author{Thomas Filk\thanks{Financial support has been provided by the 
European Research Council under the European CommunityÕs Seventh Framework 
Programme (FP7/20072013)/ ERC grant agreement no [294332], EvoEvo project.}\\
Institute of Physics, Albrecht-Ludwigs University Freiburg, Germany\\
Parmenides Foundation for the Study of Thinking, Munich, Germany}

\maketitle

\begin{abstract}
Several mechanical realizations of PR-Boxes
are discussed. Apart from superluminal correlations,
which are known not to be realizable by classical
devices, the requirements (1) PR-condition on input-output
relations, (2) non-signaling, and (3) allowing for independent
measurements are discussed separately. The examples show
that internal signaling (inside the box) does not invariably imply
the possibility of external signaling and may externally
be attributed to contextuality. The proposed model requires
two classical bits to be exchanged internally in order to realize
the defined requirements. However, with EPR correlations this
can be reduced to one classical bit.
\end{abstract}

\section{Introduction}
In a seminal article, Popescu and Rohrlich \cite{PR} 
have shown in 1992 that a maximal violation of the 
CHSH inequality (for CHSH inequalities see, e.g., 
\cite{CHSH, Clauser_Shimony}) is possible with a simple device which
became known as a PR-box (for a recent review
see \cite{Popescu}). Essentially, a PR-box is a black-box
device on which two persons -- Alice and Bob -- can make
local measurements with binary variables as input (let
$x$ be the binary input of Alice and $y$ the binary input of
Bob) and binary outputs $a$ (at Alice's side) and $b$ (at
Bob's side), such that
\begin{equation}
\label{eq_1}
              a + b = x \cdot y ~~  \mbox{mod\,}2 \, .
\end{equation}
This implies that unless $x$ and $y$ are both equal to
1, the outputs $a$ and $b$ are equal (either both 0 or
both 1), and if $x=y=1$ the outputs $a$ and $b$ are
different. 

The fact that quantum physics is more nonlocal than
classical physics but not as nonlocal as a PR-Box
(this has sometimes been called superstrong nonlocality) led
to a completely new approach of axiomatizing quantum
theory (see, e.g., \cite{Barrett,Pawlowski}. The main question in
this approach is: Can we define quantum theory by 
the degree to which non-local correlations can be
used to reduce certain communication problems. 
Some of the particular interesting consequences of 
superstrong non-locality, like e.g.\ the reduction of the
``inner product problem'' have been analyzed in \cite{Dam}. 

In addition to the property expressed by eq.\ (\ref{eq_1}) 
one generally requires that
the output of the PR-box cannot be used for signaling
between Alice and Bob. This implies that the probability
for output $b$ being 1 or 0 should not depend on the
input $x$ of Alice and, vice versa, the probability distribution
for $a$ should not depend on the input of Bob. 
Let $P(a|x,y)$ be the conditional probability for $a$
given $x$ and $y$, and similarly $P(b|x,y)$ the 
conditional probability for $b$ given $x$ and $y$, then
this implies 
\begin{equation}
    P(a|x,y)= P(a|x) ~~~{\rm and} ~~~P(b|x,y) = P(b|y) \, .
\end{equation}
This condition has also been called ``marginal
selectivity'' \cite{Ehtibar}. It implies non-signaling and
is sometimes considered as a necessary condition
for a system to be consistent with special relativity. If this
condition is violated and signaling possible, this signaling
should be bound to local causality, i.e., it should not
propagate faster than the speed of light.

As a third condition one often requires that
Alice and Bob can perform
their measurements independently in the sense that both
receive their output immediately after they entered their
input, i.e., neither of them has to wait for the other
to perform his or her experiment.

One natural question with respect to such a device is
whether it can exist or not. It can be shown that
with such a device one can violate the so-called
CHSH inequality maximally 
(see, e.g.\ \cite{Dam,Atmanspacher,Filk}).
Therefore, a device based on classical
laws of physics cannot exist, and even quantum
physics with it's non-localities cannot realize a maximal
violation of the CHSH expression. However, if we 
allow for an
internal exchange of information (and thus give up
the possibility that Alice and Bob perform their experiments
within each others causal complements of the 
light cones corresponding to the events of measurements),
one can show that such a device can be realized
by a classical electronic circuit. It is important to notice, however,
that such a device, despite the fact that internally an exchange
of information is happening, cannot necessarily be used
for signaling externally.

The question whether such a device can be built or not
may be of relevance as it has been suggested
that a PR-Box may be a logical contradiction, i.e.\ cannot
exist for principal logical reasons, and under quite
general conditions this can even be proven (\cite{Uzan}). In addition,
in this letter I will raise the question, how many bits
have to be exchanged internally within the PR-box
in order to realize such a system. 
The device introduced in this letter shows that 
internally the exchange of two classical bits 
is sufficient. However, without giving a proof (which I don't
have), it suggests that this may also be neccesarry in order to 
fulfill the no-signaling condition. Furthermore, if we replace
the random number generator used in this device by 
EPR measurements, this may be reduced to one classical
bit.

In the next section \ref{sec_PR}, I will
introduce three types of PR-boxes of increasing complexity
and in section \ref{sec_discussion}, I will discuss the
internal signaling of these PR-boxes. 
In particular, in this section I will 
discuss the number of bits which need to be exchanged
internally. A brief summary ends the article.

\section{Signaling an non-signaling PR-boxes}
\label{sec_PR}

In this section, I will construct three differen PR-boxes such 
that the final one will come as close as possible
to the abstract notion of a PR-box.
\begin{enumerate}
\item
The first PR-box will satisfy eq.\ (\ref{eq_1}), but it
can be used for signaling from Bob to Alice.
\item
The second PR-box will also satisfy eq.\ (\ref{eq_1}),
however, external signaling is no longer possible. This
PR-box is asymmetric in the sense that Bob can
perform his measurements independent of Alice,
but Alice has to wait for her result until also
Bob has performed his measurement.
\item
Finally, I construct a symmetric box which
satisfies eq.\ (\ref{eq_1}), which cannot be used for
external signaling, and both partners can perform their
measurements independently and obtain an
immediate output on their respective sides.  
\end{enumerate}
As mentioned, the second and the third PR-box cannot
be used for signaling. Of course, this does not imply
that there are no signals exchanged internally.
Therefore, I distinguish between external signaling,
which refers to the possibility of Alice and Bob to
use this device for an exchange of information, and internal signaling
which is happening inside the box but may not
be utilized by Alice and Bob.

\subsection{The signaling PR-box}

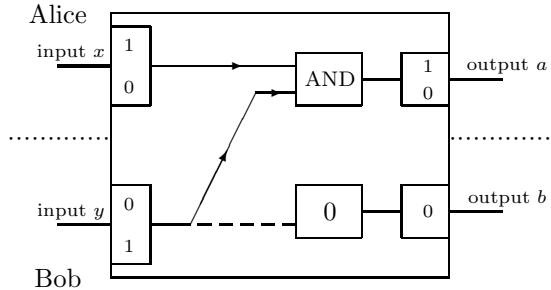
\begin{figure}[h]
\begin{picture}(190,110)(0,0)
% Boxen mit +- 1 (rechts)
\put(150,15){\line(1,0){18}}
\put(150,15){\line(0,1){20}}
\put(150,35){\line(1,0){18}}
\put(150,65){\line(1,0){18}}
\put(150,65){\line(0,1){20}}
\put(150,85){\line(1,0){18}}
%\put(160,30){\makebox(0,0){$\scriptstyle 1$}}
\put(160,25){\makebox(0,0){$\scriptstyle 0$}}
\put(160,70){\makebox(0,0){$\scriptstyle 0$}}
\put(160,80){\makebox(0,0){$\scriptstyle 1 $}}
% Box au§en
\put(40,0){\line(1,0){128}}
\put(40,0){\line(0,1){100}}
\put(40,100){\line(1,0){128}}
\put(168,0){\line(0,1){100}}
\put(47,12){\makebox(0,0){${\scriptstyle 1}$}}
\put(47,28){\makebox(0,0){${\scriptstyle 0}$}}
\put(40,5){\line(1,0){15}}
\put(40,35){\line(1,0){15}}
\put(55,5){\line(0,1){30}}
\put(40,65){\line(1,0){15}}
\put(40,95){\line(1,0){15}}
\put(55,65){\line(0,1){30}}
\put(47,72){\makebox(0,0){${\scriptstyle 0}$}}
\put(47,88){\makebox(0,0){${\scriptstyle 1}$}}
\put(25,25){\makebox(0,0){\scriptsize input $y$}}
\put(25,85){\makebox(0,0){\scriptsize input $x$}}
\put(190,30){\makebox(0,0){\scriptsize output $b$}}
\put(190,80){\makebox(0,0){\scriptsize output $a$}}

\thicklines
\put(20,20){\line(1,0){20}}
\put(20,80){\line(1,0){20}}
\put(135,25){\line(1,0){15}}
\put(135,75){\line(1,0){15}}
\put(168,25){\line(1,0){20}}
\put(168,75){\line(1,0){20}}
\thinlines
\multiput(73,20)(8,0){5}{\line(1,0){5}}
% Interne Linien und Vektoren (links)
\put(55,20){\line(1,0){15}}
\put(70,20){\line(1,2){25}}
\put(74,28){\vector(1,2){10}}
\put(95,70){\line(1,0){15}}
\put(95,70){\vector(1,0){10}}
\put(80,80){\vector(1,0){10}}
\put(55,80){\line(1,0){55}}
% Box mit Theta-Funktion
\put(110,65){\line(1,0){25}}
\put(110,65){\line(0,1){20}}
\put(110,85){\line(1,0){25}}
\put(135,65){\line(0,1){20}}
\put(123,75){\makebox(0,0){\footnotesize AND}}
%\put(117,77){\line(1,0){5}}
%\put(122,77){\line(0,1){5}}
%\put(122,82){\line(1,0){5}}
%  Box mit +1
\put(110,15){\line(1,0){25}}
\put(110,15){\line(0,1){20}}
\put(110,35){\line(1,0){25}}
\put(135,15){\line(0,1){20}}
\put(123,25){\makebox(0,0){0}}
\multiput(1,50)(3,0){13}{$\cdot$}
\multiput(168,50)(3,0){13}{$\cdot$}
\put(20,0){\makebox(0,0){Bob}}
\put(20,100){\makebox(0,0){Alice}}
\end{picture}
\caption{\label{fig_net1}%
A simple circuit which gives the output of
a PR-box. The dashed line does not carry bits but
just triggers a device which generates a 0.
This circuit can be used for signaling from Bob
to Alice.}
\end{figure}

Figure \ref{fig_net1} shows a simple circuit which
realizes equation \ref{eq_1}. The output on Bob's
side is always 0, independent of his input.
His input just triggers a device which generates
0 for his output. 

The output on Alice's side depends on both 
inputs:
only when Alice and Bob ``throw in'' a 1, the output will be 1 
on her side, in all other cases it will be 0. The output on
her side is realized by a simple logical AND-gate which leads
to eq.\ (\ref{eq_1}). While Bob can obtain his output
immediately after he inserted his input, Alice has
to wait for Bob to enter his input, otherwise the
AND-gate cannot produce an answer.

Even though eq.\ \ref{eq_1} is realized, this
box can be used for signaling, at least from Bob
to Alice. If Alice always uses 
$x=1$ as input, then obviously her output will
be exactly the same as Bob's input: 0 if Bob uses
0 as input and 1 if Bob uses 1 as input.
There is a strict correlation between the output
on Alice's side with the input on Bob's side.

Let us finally remark that in this device internally 
one bit of information
is signaled from Bob to Alice.

\subsection{A non-signaling PR-box}

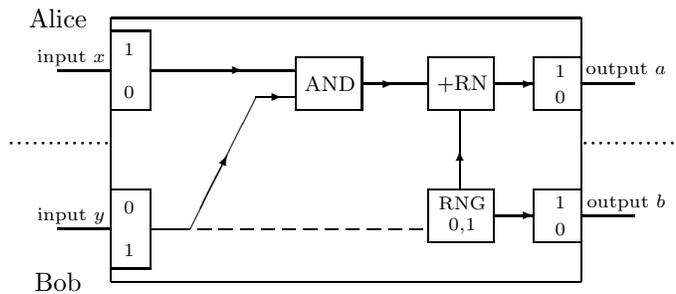
\begin{figure}[h]
\begin{picture}(240,110)(0,0)
% Boxen mit +- 1 (rechts)
\put(200,15){\line(1,0){18}}
\put(200,15){\line(0,1){20}}
\put(200,35){\line(1,0){18}}
\put(200,65){\line(1,0){18}}
\put(200,65){\line(0,1){20}}
\put(200,85){\line(1,0){18}}
\put(210,30){\makebox(0,0){$\scriptstyle 1$}}
\put(210,20){\makebox(0,0){$\scriptstyle 0$}}
\put(210,70){\makebox(0,0){$\scriptstyle 0$}}
\put(210,80){\makebox(0,0){$\scriptstyle 1 $}}
% Box au§en
\put(40,0){\line(1,0){178}}
\put(40,0){\line(0,1){100}}
\put(40,100){\line(1,0){178}}
\put(218,0){\line(0,1){100}}
\put(47,12){\makebox(0,0){${\scriptstyle 1}$}}
\put(47,28){\makebox(0,0){${\scriptstyle 0}$}}
\put(40,5){\line(1,0){15}}
\put(40,35){\line(1,0){15}}
\put(55,5){\line(0,1){30}}
\put(40,65){\line(1,0){15}}
\put(40,95){\line(1,0){15}}
\put(55,65){\line(0,1){30}}
\put(47,72){\makebox(0,0){${\scriptstyle 0}$}}
\put(47,88){\makebox(0,0){${\scriptstyle 1}$}}
%
%\put(115,25){\line(1,0){85}}
\put(135,75){\line(1,0){25}}
\put(172,35){\line(0,1){30}}
\put(185,75){\vector(1,0){15}}
\put(185,25){\vector(1,0){15}}
\put(172,35){\vector(0,1){15}}
\put(135,75){\vector(1,0){12}}

\put(25,25){\makebox(0,0){\scriptsize input $y$}}
\put(25,85){\makebox(0,0){\scriptsize input $x$}}
\put(235,30){\makebox(0,0){\scriptsize output $b$}}
\put(235,80){\makebox(0,0){\scriptsize output $a$}}

\thicklines
\put(20,20){\line(1,0){20}}
\put(20,80){\line(1,0){20}}
\put(218,25){\line(1,0){20}}
\put(218,75){\line(1,0){20}}
\thinlines
%
% Interne Linien und Vektoren (links)
\multiput(73,20)(8,0){11}{\line(1,0){5}}
\put(55,20){\line(1,0){15}}
\put(70,20){\line(1,2){25}}
\put(74,28){\vector(1,2){10}}
\put(95,70){\line(1,0){15}}
\put(95,70){\vector(1,0){10}}
\put(80,80){\vector(1,0){10}}
\put(55,80){\line(1,0){55}}
% Box mit Theta-Funktion
\put(110,65){\line(1,0){25}}
\put(110,65){\line(0,1){20}}
\put(110,85){\line(1,0){25}}
\put(135,65){\line(0,1){20}}
\put(123,75){\makebox(0,0){\footnotesize AND}}
%\put(117,77){\line(1,0){5}}
%\put(122,77){\line(0,1){5}}
%\put(122,82){\line(1,0){5}}
% Box nur Negation
\put(160,65){\line(1,0){25}}
\put(160,65){\line(0,1){20}}
\put(160,85){\line(1,0){25}}
\put(185,65){\line(0,1){20}}
\put(173,75){\makebox(0,0){\footnotesize $+$RN}}
%  Box mit RNG
\put(160,15){\line(1,0){25}}
\put(160,15){\line(0,1){20}}
\put(160,35){\line(1,0){25}}
\put(185,15){\line(0,1){20}}
\put(173,30){\makebox(0,0){{\scriptsize RNG}}}
\put(173,21){\makebox(0,0){${\scriptstyle 0,1}$}}
\multiput(1,50)(3,0){13}{$\cdot$}
\multiput(218,50)(3,0){13}{$\cdot$}
\put(20,0){\makebox(0,0){Bob}}
\put(20,100){\makebox(0,0){Alice}}
\end{picture}
\caption{\label{fig_net2}%
The output of this circuit also satisfies
eq.\ \ref{eq_1}. However, due to the random number
generator, the outputs are not related to the inputs
and cannot be used for signaling. The dashed line
on Bob's side now represents a trigger for
a random number generator. The output on Bob's side
does not depend on his input but equals the 
generated random number.}
\end{figure}

Now let us consider Fig.\ \ref{fig_net2}. This
circuit differes from the previous one in that
the device which simply generated a 0 on
Bob's side has now been replaced by
a random number generator (RNG). This random
number generator not only determines the
output on Bob's side (which now is random)
but it also adds this random number to the 
output of the AND-gate on Alice's side: 
essentially, it switches the output if
the random number is 1 and it leaves the
output unchanged if the random number is 0.
This is happening in the ``$+$RN'' (``add random
number'')-device.

Obviously, the outputs on both sides are
always equal, unless both use 1 as an input,
in which case the outputs are different. Therefore,
this device still satisfies eq.\ (\ref{eq_1}). However,
due to the random number generator, this
box cannot be used for external signaling,
as statistically the output on Alice's side is no longer
correlated with the input on Bob's side.

Note that now two bits are transmitted
internally from Bob's side to Alice's side:
one bit corresponds to the input of Bob
and the second one to the result of the
random number generator. 

This PR-box is still asymmetric in the sense that
Bob can perform his measurement and obtains
an immediate output while Alice has to wait
for Bob's input. In order to become a PR-box in
the usual sense (except, as mentioned before,
for the impossibility of superluminuous 
internal signaling), Alice and Bob should be able
to perform their measurements independently
and obtain immediate outputs.

\subsection{A PR-box with immediate outputs}

\begin{figure}[htb]
\begin{picture}(240,170)(10,-5)
% Box au§en
\put(40,0){\line(1,0){228}}
\put(40,0){\line(0,1){150}}
\put(40,150){\line(1,0){228}}
\put(268,0){\line(0,1){150}}
% Boxen mit 1 0  (rechts)
\put(250,15){\line(1,0){18}}
\put(250,15){\line(0,1){20}}
\put(250,35){\line(1,0){18}}
\put(250,115){\line(1,0){18}}
\put(250,115){\line(0,1){20}}
\put(250,135){\line(1,0){18}}
\put(260,30){\makebox(0,0){$\scriptstyle 1$}}
\put(260,20){\makebox(0,0){$\scriptstyle 0$}}
\put(260,120){\makebox(0,0){$\scriptstyle 0$}}
\put(260,130){\makebox(0,0){$\scriptstyle 1 $}}
% Boxen mit 1 0 links
\put(47,12){\makebox(0,0){${\scriptstyle 1}$}}
\put(47,28){\makebox(0,0){${\scriptstyle 0}$}}
\put(40,5){\line(1,0){15}}
\put(40,35){\line(1,0){15}}
\put(55,5){\line(0,1){30}}
\put(40,115){\line(1,0){15}}
\put(40,145){\line(1,0){15}}
\put(55,115){\line(0,1){30}}
\put(47,122){\makebox(0,0){${\scriptstyle 0}$}}
\put(47,138){\makebox(0,0){${\scriptstyle 1}$}}
\put(25,25){\makebox(0,0){\scriptsize input $y$}}
\put(25,135){\makebox(0,0){\scriptsize input $x$}}
\put(285,30){\makebox(0,0){\scriptsize output $b$}}
\put(285,130){\makebox(0,0){\scriptsize output $a$}}
\thicklines
\put(20,20){\line(1,0){20}}
\put(20,130){\line(1,0){20}}
\put(268,25){\line(1,0){20}}
\put(268,125){\line(1,0){20}}
\thinlines
%  Store + AND + +RN  oben
\put(90,120){\line(1,0){40}}
\put(90,120){\line(0,1){20}}
\put(90,140){\line(1,0){40}}
\put(130,120){\line(0,1){20}}
\put(110,130){\makebox(0,0){\footnotesize STORE}}
\put(150,120){\line(1,0){30}}
\put(150,120){\line(0,1){20}}
\put(150,140){\line(1,0){30}}
\put(180,120){\line(0,1){20}}
\put(165,130){\makebox(0,0){\footnotesize AND}}
\put(200,120){\line(1,0){30}}
\put(200,120){\line(0,1){20}}
\put(200,140){\line(1,0){30}}
\put(230,120){\line(0,1){20}}
\put(215,130){\makebox(0,0){\footnotesize +RN}}
\put(170,90){\line(1,0){30}}
\put(170,90){\line(0,1){20}}
\put(170,110){\line(1,0){30}}
\put(200,90){\line(0,1){20}}
\put(185,100){\makebox(0,0){\footnotesize RNG}}
%  Store + AND + +RN  unten
\put(90,10){\line(1,0){40}}
\put(90,10){\line(0,1){20}}
\put(90,30){\line(1,0){40}}
\put(130,10){\line(0,1){20}}
\put(110,20){\makebox(0,0){\footnotesize STORE}}
\put(150,10){\line(1,0){30}}
\put(150,10){\line(0,1){20}}
\put(150,30){\line(1,0){30}}
\put(180,10){\line(0,1){20}}
\put(165,20){\makebox(0,0){\footnotesize AND}}
\put(200,10){\line(1,0){30}}
\put(200,10){\line(0,1){20}}
\put(200,30){\line(1,0){30}}
\put(230,10){\line(0,1){20}}
\put(215,20){\makebox(0,0){\footnotesize +RN}}
\put(170,40){\line(1,0){30}}
\put(170,40){\line(0,1){20}}
\put(170,60){\line(1,0){30}}
\put(200,40){\line(0,1){20}}
\put(185,50){\makebox(0,0){\footnotesize RNG}}
% Vektoren und Linien oben
\put(55,133){\line(1,0){35}}
\put(55,133){\vector(1,0){20}}
\put(130,133){\line(1,0){20}}
\put(130,133){\vector(1,0){13}}
\put(130,127){\line(1,0){20}}
\put(130,127){\vector(1,0){13}}
\put(180,130){\line(1,0){20}}
\put(180,130){\vector(1,0){13}}
\put(210,100){\line(0,1){20}}
\put(210,100){\vector(0,1){10}}
\put(210,100){\line(2,1){40}}
\put(210,100){\vector(2,1){20}}
\put(210,100){\line(0,-1){70}}
\put(210,102){\vector(0,-1){40}}
\put(200,100){\line(1,0){10}}
\put(200,100){\vector(1,0){10}}
\put(230,130){\line(4,-1){20}}
\put(230,130){\vector(4,-1){10}}
\multiput(109,115)(0,-8){3}{\line(0,1){5}}
\multiput(109,99)(8,0){8}{\line(1,0){5}}
% Vektoren und Linien unten
\put(55,17){\line(1,0){35}}
\put(55,17){\vector(1,0){20}}
\put(130,17){\line(1,0){20}}
\put(130,17){\vector(1,0){13}}
\put(130,23){\line(1,0){20}}
\put(130,23){\vector(1,0){13}}
\put(180,20){\line(1,0){20}}
\put(180,20){\vector(1,0){13}}
\put(220,30){\line(0,1){20}}
\put(220,50){\vector(0,1){30}}
\put(220,50){\line(0,1){70}}
\put(220,50){\vector(0,-1){10}}
\put(200,50){\line(1,0){20}}
\put(208,50){\vector(1,0){10}}
\put(220,50){\line(3,-2){30}}
\put(220,50){\vector(3,-2){15}}
\put(230,20){\line(4,1){20}}
\put(230,20){\vector(4,1){10}}
\multiput(109,30)(0,8){3}{\line(0,1){5}}
\multiput(109,52)(8,0){8}{\line(1,0){5}}
% Vektoren und Linien kreuzend
\put(58,133){\line(1,-4){27}}
\put(58,133){\vector(1,-4){12}}
\put(85,25){\line(1,0){5}}
\put(58,17){\line(1,4){27}}
\put(58,17){\vector(1,4){12}}
\put(85,125){\line(1,0){5}}
\multiput(1,75)(3,0){13}{$\cdot$}
\multiput(268,75)(3,0){13}{$\cdot$}
\put(20,0){\makebox(0,0){Bob}}
\put(20,150){\makebox(0,0){Alice}}
\end{picture}
\caption{\label{fig_net3}%
This PR-box yields immediate results when Alice and Bob
perform their measurements. As before, dashed lines just
trigger the random number generator RNG; 
they do not carry bits related to input or output. For details
see main text.}
\end{figure}
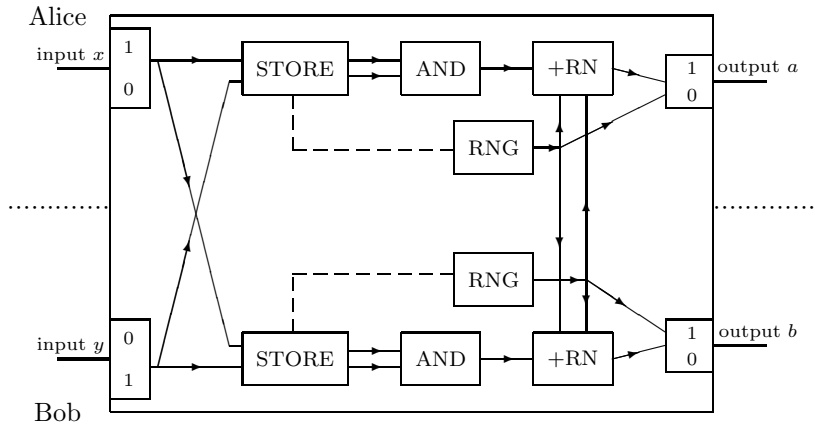

Figure \ref{fig_net3} shows a device which
satisfies eq.\ (\ref{eq_1}), cannot be used for
(external) signaling and allows both sides
to perform their measurements with immediate
results, independent of what happens at the other side. 
This is achieved by ``symmetrizing'' the 
box of the last section: whoever makes the
first measurement triggers the random number
generator and obtains the result of the random
number generator as an output. 

Let us consider the device in more detail. The device is symmetric,
but in order to be specific we assume that Alice makes the first 
measurement by inserting the input $x$ into the box. This input
$x$ is sent to her STORE-gate which simply stores this input 
until also the second input $y$ from Bob has entered. Furthermore,
$x$ is transmitted to Bob and stored in his STORE-gate. The
STORE-gates ``know'' who has made the first measurement. 
For Alice, the STORE-gate blocks the further transmission 
of the input but instead triggers a random number generator RNG
which produces a signal and sends this signal to Alice's output.
She gets her immediate result. Furthermore, the random number
generator sends it's result to the ``+RN''-device on Bob's side which
is stored there until Bob makes his measurement.

On Bob's side the STORE-gate
receives Alice's input and now blocks the random number
generator. Instead it waits for Bob's input $y$ (the second
measurement). The result is transfered to the AND-gate which
gives an output 1 if both $x$ and $y$ had been 1, otherwise
it yields the output 0. The AND-gates transfers the result to
Bob's ``+RN''-device where the random number which
originally came from Alice is added. The result is transfered
to Bob's output who gets his result immediately.

Note that dashed lines in fig.\ \ref{fig_net3} 
do not transmit any bits related to the inputs
or outputs but only serve to trigger the random
number generators. Furthermore,
Alice and Bob should not perform their measurements within
their light-cone complements as in this case both STORE-gates
will ``assume'' that the measurement from the nearer side
has been performed first and trigger the random number
generators. In this case the outputs are random and
do not satisfy the PR-box condition. 

\section{Reducing the internal exchange bits}
\label{sec_discussion}

PR-boxes are not meant to be built. However, it
is instructive to see to which extend such a device
can be realized as a classical physical system. 
The explicit realization -- bound, however, to the
constraints of special relativity -- proves that such
a box is not logically inconsistent.

The more interesting question is related to the
number of bits which have to be exchanged 
internally between the two sides. As was to be
expected, only one classical bit had to be exchanged
in order to satisfy eq.\ \ref{eq_1}, as this equation
can be realized by a simple AND-gate. However,
in order to make the device non-signalling (externally),
it seems that at least one more classical bit has
to be exchanged internally, which in the discussed
case was the result of the random number generator. 

This raises the natural question, whether the number
of exchanged bits can be reduced if we use
quantum correlations. Of course, it is known from
superdense coding \cite{Bennett_Wiesner} that
two classical bits can be exchanged by transmitting
only one photon (which has an entangled partner on the
other side). However, we also may replace the random
number generator by an EPR state thus reducing the
number of transferred classical bits to one.

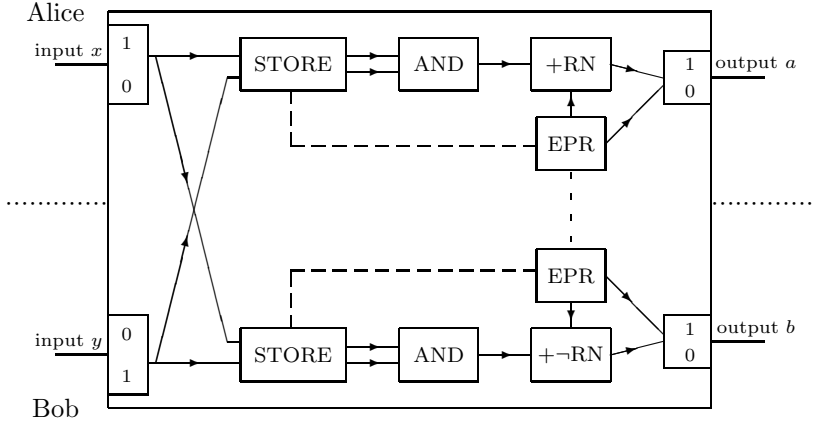
\begin{figure}[htb]
\begin{picture}(240,170)(10,-5)
% Box au§en
\put(40,0){\line(1,0){228}}
\put(40,0){\line(0,1){150}}
\put(40,150){\line(1,0){228}}
\put(268,0){\line(0,1){150}}
% Boxen mit 1 0  (rechts)
\put(250,15){\line(1,0){18}}
\put(250,15){\line(0,1){20}}
\put(250,35){\line(1,0){18}}
\put(250,115){\line(1,0){18}}
\put(250,115){\line(0,1){20}}
\put(250,135){\line(1,0){18}}
\put(260,30){\makebox(0,0){$\scriptstyle 1$}}
\put(260,20){\makebox(0,0){$\scriptstyle 0$}}
\put(260,120){\makebox(0,0){$\scriptstyle 0$}}
\put(260,130){\makebox(0,0){$\scriptstyle 1 $}}
% Boxen mit 1 0 links
\put(47,12){\makebox(0,0){${\scriptstyle 1}$}}
\put(47,28){\makebox(0,0){${\scriptstyle 0}$}}
\put(40,5){\line(1,0){15}}
\put(40,35){\line(1,0){15}}
\put(55,5){\line(0,1){30}}
\put(40,115){\line(1,0){15}}
\put(40,145){\line(1,0){15}}
\put(55,115){\line(0,1){30}}
\put(47,122){\makebox(0,0){${\scriptstyle 0}$}}
\put(47,138){\makebox(0,0){${\scriptstyle 1}$}}
\put(25,25){\makebox(0,0){\scriptsize input $y$}}
\put(25,135){\makebox(0,0){\scriptsize input $x$}}
\put(285,30){\makebox(0,0){\scriptsize output $b$}}
\put(285,130){\makebox(0,0){\scriptsize output $a$}}
\thicklines
\put(20,20){\line(1,0){20}}
\put(20,130){\line(1,0){20}}
\put(268,25){\line(1,0){20}}
\put(268,125){\line(1,0){20}}
\thinlines
%  Store + AND + +RN  oben
\put(90,120){\line(1,0){40}}
\put(90,120){\line(0,1){20}}
\put(90,140){\line(1,0){40}}
\put(130,120){\line(0,1){20}}
\put(110,130){\makebox(0,0){\footnotesize STORE}}
\put(150,120){\line(1,0){30}}
\put(150,120){\line(0,1){20}}
\put(150,140){\line(1,0){30}}
\put(180,120){\line(0,1){20}}
\put(165,130){\makebox(0,0){\footnotesize AND}}
\put(200,120){\line(1,0){30}}
\put(200,120){\line(0,1){20}}
\put(200,140){\line(1,0){30}}
\put(230,120){\line(0,1){20}}
\put(215,130){\makebox(0,0){\footnotesize +RN}}
\put(202,90){\line(1,0){26}}
\put(202,90){\line(0,1){20}}
\put(202,110){\line(1,0){26}}
\put(228,90){\line(0,1){20}}
\put(215,100){\makebox(0,0){\footnotesize EPR}}
%  Store + AND + +RN  unten
\put(90,10){\line(1,0){40}}
\put(90,10){\line(0,1){20}}
\put(90,30){\line(1,0){40}}
\put(130,10){\line(0,1){20}}
\put(110,20){\makebox(0,0){\footnotesize STORE}}
\put(150,10){\line(1,0){30}}
\put(150,10){\line(0,1){20}}
\put(150,30){\line(1,0){30}}
\put(180,10){\line(0,1){20}}
\put(165,20){\makebox(0,0){\footnotesize AND}}
\put(200,10){\line(1,0){30}}
\put(200,10){\line(0,1){20}}
\put(200,30){\line(1,0){30}}
\put(230,10){\line(0,1){20}}
\put(215,20){\makebox(0,0){\footnotesize +$\neg$RN}}
\put(202,40){\line(1,0){26}}
\put(202,40){\line(0,1){20}}
\put(202,60){\line(1,0){26}}
\put(228,40){\line(0,1){20}}
\put(215,50){\makebox(0,0){\footnotesize EPR}}
\multiput(215,63)(0,8){4}{\line(0,1){2}}
% Vektoren und Linien oben
\put(55,133){\line(1,0){35}}
\put(55,133){\vector(1,0){20}}
\put(130,133){\line(1,0){20}}
\put(130,133){\vector(1,0){13}}
\put(130,127){\line(1,0){20}}
\put(130,127){\vector(1,0){13}}
\put(180,130){\line(1,0){20}}
\put(180,130){\vector(1,0){13}}
\put(215,110){\line(0,1){10}}
\put(215,118){\vector(0,1){0}}
\put(228,100){\line(1,1){22}}
\put(228,100){\vector(1,1){10}}
%\put(210,100){\line(0,-1){70}}
%\put(210,102){\vector(0,-1){40}}
%\put(200,100){\line(1,0){10}}
%\put(200,100){\vector(1,0){10}}
\put(230,130){\line(4,-1){20}}
\put(230,130){\vector(4,-1){10}}
\multiput(109,115)(0,-8){3}{\line(0,1){5}}
\multiput(109,99)(8,0){12}{\line(1,0){5}}
% Vektoren und Linien unten
\put(55,17){\line(1,0){35}}
\put(55,17){\vector(1,0){20}}
\put(130,17){\line(1,0){20}}
\put(130,17){\vector(1,0){13}}
\put(130,23){\line(1,0){20}}
\put(130,23){\vector(1,0){13}}
\put(180,20){\line(1,0){20}}
\put(180,20){\vector(1,0){13}}
\put(215,30){\line(0,1){10}}
%\put(220,50){\vector(0,1){30}}
%\put(220,50){\line(0,1){70}}
\put(215,33){\vector(0,-1){0}}
%\put(200,50){\line(1,0){20}}
%\put(208,50){\vector(1,0){10}}
\put(228,50){\line(1,-1){22}}
\put(228,50){\vector(1,-1){10}}
\put(230,20){\line(4,1){20}}
\put(230,20){\vector(4,1){10}}
\multiput(109,30)(0,8){3}{\line(0,1){5}}
\multiput(109,52)(8,0){12}{\line(1,0){5}}
% Vektoren und Linien kreuzend
\put(58,133){\line(1,-4){27}}
\put(58,133){\vector(1,-4){12}}
\put(85,25){\line(1,0){5}}
\put(58,17){\line(1,4){27}}
\put(58,17){\vector(1,4){12}}
\put(85,125){\line(1,0){5}}
\multiput(1,75)(3,0){13}{$\cdot$}
\multiput(268,75)(3,0){13}{$\cdot$}
\put(20,0){\makebox(0,0){Bob}}
\put(20,150){\makebox(0,0){Alice}}
\end{picture}
\caption{\label{fig_net4}%
The random number generator of Fig.\ \ref{fig_net3} has
been replaced by an EPR-Bohm state of entangled particles 
shared by Alice and Bob. Internal measurements on
this state can create correlated random numbers on Alice's and
Bob's side without the exchange of information (because of the
anti-correlation in EPR-states, Bob adds the negation of his
EPR-result to the output of the AND-gate).}
\end{figure}

The set-up would be as follows (Fig.\ \ref{fig_net4}): 
Both Alice and Bob
share an entangled EPR state (e.g.\ realized by photons).
Whoever makes the first measurement (say, Alice) by 
inserting her input bit $x$, triggers several events:
first, a classical signal is sent to Bob transmitting the
input of Alice. This signal is stored in the AND-gate of
Bob until Bob makes his measurement. Furthermore,
the input of $x$ triggers on Alice's side an EPR measurement 
on one of the entangled photons with respect to
a predefined basis. The result ($+1$ or $-1$ translated
into $1$ and $0$, respectively) will be the result
of Alice's measurement ($a$). When the second measurement
is performed (by Bob, inserting his input $y$) internally
two things happen: first, the input $y$ is send to Bob's
AND-gate and processed with Alice's input $x$ which is
already there. Furthermore, on Bob's side a measurement
is made on his EPR photon with respect to the same
basis as Alice's measurement. If Alice and Bob have
initially agreed to interprete the result of the EPR
measurement ($+1$ and $-1$) as $1$ 
and 0, Bob has to reverse this result (because of the anti-correlations
in the EPR states) before adding it to the output of the
AND-gate in order to fulfill the PR-requirement.
Hence, the EPR correlation reduces the two classical
bits which have to be exchanged to one classical
bit. In general, EPR correlations can be used to exchange
random numbers without the exchange of classical bits.

Once more it should be emphasized that, although
both sides can perform their measurements independently
and obtain an immediate answer, the device does
not work when both sides perform their measurements
simultaneously within the causal complements of the light cones
of each others measurements. In this case both sides trigger
their random number generators and the
answers will not satisfy the PR-box condition (eq.\ \ref{eq_1}).
It is known that a classical device with this property
cannot exist. 

\section{Summary and Conclusion}

I investigated several types of ``real'' PR-boxes of increasing
complexity. Apart from measurements performed within
causal complements, the constructed classical PR-box 
satisfies (1) the PR condition, (2) the no-signaling 
property, and (3) the possibility for independent 
measurements with immediate answers.

Classically it seems that in order to guarantee point (1) and (2),
two classical bits have to be exchanged, of which the
first one is an exchange of one of the inputs and the
second one an exchange of a random number. The
second classical bit can be replaced by a
``non-signaling'' EPR-measurement.

It has been shown that although
there is internal signaling between the sides of Alice and Bob, 
the device cannot be used for external signaling. A
violation of Bell-type inequalities together with the
non-signaling property has been
coined contextuality (see, e.g., \cite{Ehtibar}). So, what
externally looks like contextuality may internally be
due to signaling.

\section*{Acknowledgements}

I acknowledge several stimulating discussions during the
QI 2105 conference in Filzbach, Switzerland. I am 
particularly grateful to Alexei Grinbaum and Pierre Uzan
for their views on PR-boxes.

\end{document}